\documentclass{article}

\usepackage{arxiv}
\usepackage[utf8]{inputenc} 
\usepackage[T1]{fontenc}    
\usepackage{hyperref}       
\usepackage{url}            
\usepackage{booktabs}       
\usepackage{amsfonts}       
\usepackage{nicefrac}       
\usepackage{microtype}      
\usepackage{lipsum}		
\usepackage{graphicx}
\usepackage{doi}
\usepackage{wrapfig}
\usepackage{algorithm}
\usepackage[noend]{algpseudocode}
\usepackage[bottom]{footmisc}
\usepackage{amsmath}
\usepackage{xcolor}         
\newtheorem{theorem}{Theorem}

\title{Integrate-and-Fire from a Mathematical\\and Signal Processing Perspective
\thanks{This work was supported (1) by the 'University SAL Labs' initiative of Silicon Austria Labs (SAL) and its Austrian partner universities for applied fundamental research for electronic based systems, (2) by Austrian ministries BMK, BMDW, and the State of UpperAustria in the frame of SCCH and its S3AI project, part of the COMET Programme managed by FFG.}
}

\author{ 
{
\hspace{1mm}Bernhard A.~Moser}\thanks{double affiliation: Software Competence Center Hagenberg (SCCH), 4232 Hagenberg, Austria} \\
	Institute of Signal Processing \\
	Johannes Kepler University of Linz\\
	\texttt{bernhard.moser@\{scch.at,jku.at\}} 
	\And
	{\hspace{1mm}Anna Werzi}\\
   Institute of Signal Processing \\
   Johannes Kepler University of Linz (JKU)\\
  \texttt{anna.werzi@jku.at}	
	\And
	{\hspace{1mm}Michael Lunglmayr} \\
	Institute of Signal Processing\\
	Johannes Kepler University of Linz (JKU) \\
	\texttt{michael.lunglmayr@jku.at} 
	}

\renewcommand{\shorttitle}{\textit{arXiv} Integrate-and-Fire from a Mathematical Perspective}

\hypersetup{
pdftitle={A template for the arxiv style},
pdfsubject={q-bio.NC, q-bio.QM},
pdfauthor={Bernhard A.~Moser, Michael Lunglmayr},
pdfkeywords={Leaky Integrate-and-Fire, Neuromorphic Computing, Spiking Neuron Model, Threshold-based Sampling},
}

\begin{document}
\maketitle

\begin{abstract}
Integrate-and-Fire (IF) is an idealized model of the spike-triggering mechanism of a biological neuron. It is used to realize the bio-inspired event-based principle of information processing in neuromorphic computing.
We show that IF is closely related to the concept of Send-on-Delta (SOD) as used in threshold-based sampling. 
It turns out that the IF model can be adjusted in a way that SOD can be understood as differential version of IF.
As a result, we gain insight into the underlying metric structure based on the Alexiewicz norm with consequences for clarifying the underlying signal space including bounded integrable signals with superpositions of finitely many Dirac impulses, the identification of a maximum sparsity property, error bounds for signal reconstruction and a characterization in terms of sparse regularization.
\end{abstract}


\section{Introduction}
Event-based mechanisms are fundamental in biological information processing~\cite{Tayarani-Najaran2021}.
This contributes to the realization of sparsity and efficiency and motivates engineers to look for bio-inspired approaches to sensing, signal processing and learning~\cite{bookGerstner2014}.
A key principle of neuromorphic sensing is to substitute the traditional approach of periodic sampling in favor of an event-driven scheme that imitates sampling in the nervous system, where events are primarily triggered by a change in the perceived stimulus. This type of temporal information coding is also referred to in the literature as Threshold-Based Representation (TBR)~\cite{Forno2022}, for which threshold-based mechanisms based on Integrate-and-Fire (IF) and Send-on-Delta (SOD) provide the simplest mathematical models~\cite{Miskowicz2006,Miskowicz2017}.
IF is also used to realize spiking neural networks, see e.g.~\cite{WOO2022}.
In contrast to Shannon-based sampling, which is inherently inefficient because the equidistant time interval between samples is defined by a worst-case parameter (the maximum frequency of the signal), threshold-based sampling adaptively adjusts to local signal characteristics by only sampling if a 
certain comparative characteristic of the signal changes by a certain amount, i.e., by a defined threshold value $\vartheta$. 
For IF the comparative characteristic is based on the integral, for SOD it is the amplitude.
In other words, while IF triggers a positive, respectively a negative event (also referred to as {\it spike}), depending on the condition whether the integral of the signal exceeds or undercuts the next level, SOD generates spikes by checking the condition whether the signal covers a range that exceeds the threshold value which is equivalent to level-crossing sampling with hysteresis~\cite{Zhao2022}. 
However, the principle of IF is used in different configurations, e.g. depending on how the reset of the neuron is realized after a triggering event. While the state of the art limits the application of SOD to continuous signals, we generalize it so that it can also be applied to discontinuous signals. 
This limitation prevents an efficient application to signals composed of jumps, as e.g. encounter in communication and  measurement  applications (e.g. in applications using accelerometer data as used below).
While the mathematical principles are outlined in this article, their exploitation, e.g. for a novel design approach for 
adapting analog-to-spike converters for signals with jumps

First, we clarify notation and definitions in Section~\ref{sec:math0}.
Section~\ref{sec:math1} recalls recent results on mathematical properties of IF, comprising findings about the underlying signal space~\cite{MoserLunglmayr_AIROV2024}, 
metric properties such as quasi isometry~\cite{Moser2017Similarity,MoserLunglmayr2019QuasiIsometry}, its
quantization property~\cite{MOSER2024128190} and a result on maximal sampling sparsity~\cite{moser2024samplingsparsityArxiv}.
In Section~\ref{s:Reconstruction} we discuss approaches to signal reconstruction and in Section~\ref{s:TV} we establish a connection to sparse regularization.
Examples based on acceleration data in Section~\ref{s:Examples} illustrate these findings by outlining the different behavior between IF based on {\it reset-to-mod}, also referred to as IF/mod in comparison with IF based on {\it reset-by-subtraction}, i.e., IF/sub.

\section{Mathematical Preliminaries}
\label{sec:math0}
Abstraction and idealization of biological neurons lead to 
mathematical models referred to as Integrate-and-Fire (IF).
So, spikes triggering events take effect immediately and without delay.
Depending on the precise definition of its building blocks known as triggering and re-initialization (reset) mechanisms,
we obtain different variants of IF~\cite{MOSER2024128190}. 

\begin{figure}[htbp]
	\centering		
		\includegraphics[width=0.4\textwidth]{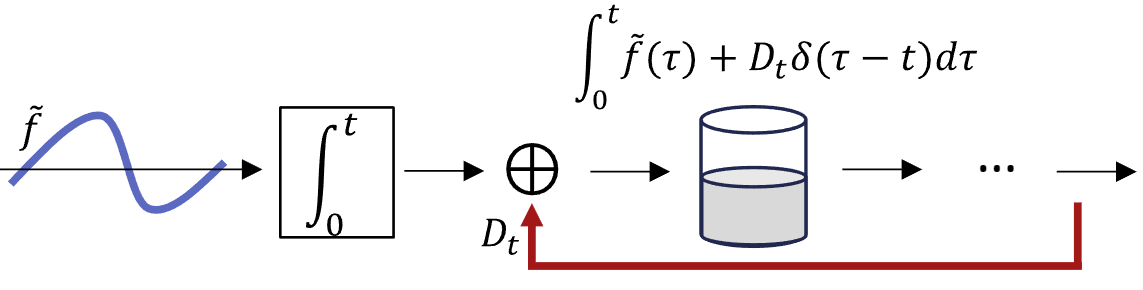}		
  		\caption{Schematic description of integration within an IF neuron with spike feedback connection leading to signals with superimposed Dirac pulses.}
	\label{fig:Dirac}
\end{figure}

\begin{figure}[htbp]
	\centering
	\includegraphics[width=0.35\linewidth]{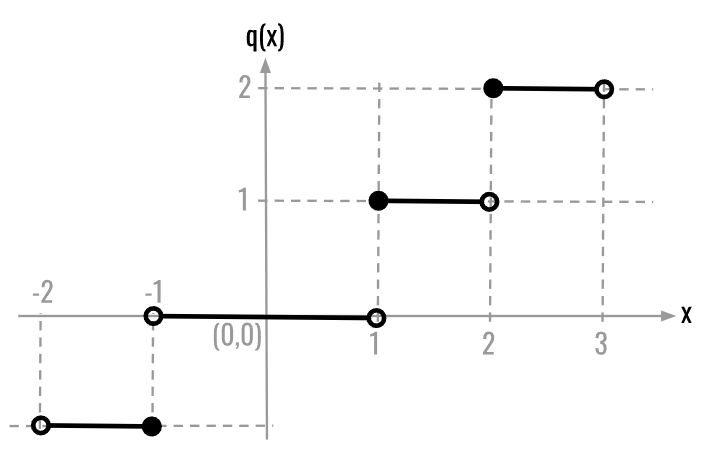}
	\caption{Standard quantization by truncation $q(x):=q_{\vartheta}(x)$ for $\vartheta = 1$, i.e., $q_{\vartheta}(x)= \vartheta q(x/\vartheta)$.}
	\label{fig:q}
\end{figure}

The integrate-and-fire (IF) operation recursively determines the time points $t_{k}$ based on  level crossings. In the context of this paper we consider positive and negative levels given by multiples of a threshold parameter 
$\vartheta>0$. 
Then IF is defined by the mapping
\begin{equation}
\label{eq:DefIF}
\mbox{IF}_{\vartheta}: \mathcal{F} \rightarrow \mathbb{S}_{\vartheta}.
\end{equation}  
This mapping converts a signal $f: [t_a, t_b]\rightarrow \mathbb{R}$ from a signal space $\mathcal{F}$ under consideration  into a spike train $\mbox{IF}_{\vartheta}(f)= s\in \mathbb{S}_{\vartheta}$ given by $s(t) = \sum_k s_k\delta(t - t_k)$.
$(t_k)_k$ is a finite sequence of increasing times $t_k$ and
$s_k \in \vartheta \mathbb{Z} = \{\vartheta\cdot k:\, k \in \mathbb{Z}\}$ denotes the amplitude of the pulse (spike) at time $t_k$. $\mathbb{S}_{\vartheta}$ denotes the vector space of all spike trains with amplitudes $s_k \in \vartheta\, \mathbb{Z}$. The time points $t_{k}$ are determined recursively by
\begin{equation}
\label{eq:IF}
t_{k+1} :=\inf\left\{t\geq t_k: \,
\left| u(t_k, t)  \right| 
\geq \vartheta\right\}, 
\end{equation}
where (the membrane potential) 
\begin{equation}
\label{eq:u}
u(t_k, t) := 
\int_{t_k}^{t} f(\tau) - r_k \delta(\tau-t_k) d\tau.
\end{equation}
The time $t_{k+1}$ is the first time after $t_{k}$ that causes the integral in \eqref{eq:u} to violate the sub-threshold condition $|u(t_k,t)| < \vartheta$.
The reset term $r_k$ in \eqref{eq:u} refers to the discharge of the membrane potential immediately after a spike has been triggered. In the context of spiking neural networks~\cite{bookGerstner2014} two variants to define $r_k$ are used. First, there is the {\it reset-to-zero} variant that 
resets the potential to zero, further there is the variant referred to as {\it reset-by-subtraction} by which the membrane potential is reduced, respectively increased by the threshold value depending on whether the membrane's potential reaches the positive or the negative threshold level $\pm \vartheta$. For continuous membrane potentials both variants are equal. For discontinuous potential curves, e.g. due to superimposed Dirac impulses resulting from spike injections from other neurons, see Fig.~\ref{fig:Dirac}, the different variants may cause different behaviors. Following the modeling principle of integrate-and-fire to idealize events by instantaneous changes, in~\cite{MOSER2024128190} 
we proposed a third variant, {\it reset-to-mod}, that instantaneously discharges the membrane potential by a multiple of the threshold value to reach subthreshold level again. 
The resulting adaption is a modulo operation that results in the signed remainder after dividing the membrane potential by the threshold value.
To sum up, we have
\begin{equation}
\label{eq:reset}
r_k := 
\left\{
\begin{array}[2]{lcl}
	U_k & \ldots & \mbox{for {\it reset-to-zero}}, \\
	\mbox{sgn}(U_k)\, \vartheta\  & \ldots & \mbox{for {\it reset-by-subtraction}},  \\
        q_{\vartheta}(U_k)    & \ldots & \mbox{for {\it reset-to-mod}},	
\end{array}
\right.
\end{equation}
where $U_k:= u(t_{k-1}, t_k)$, $\mbox{sgn}(x) \in \{-1,0,1\}$ is the signum function and 
\begin{equation}
\label{eq:q}
q_{\vartheta}(x) := \vartheta\, q(x/\vartheta) =
\mbox{sgn}(x)\max\{\vartheta k \in \vartheta\mathbb{Z}: k\vartheta \leq |x|\} 
\end{equation}
realizes a scaled variant of the standard quantization by truncation $q$, see Fig.~\ref{fig:q}.
In this paper we do not consider the variant based on {\it reset-to-zero}. For the other variants we also use the notation $\mbox{IF}_{\vartheta}^{S}$ for {\it reset-by-subtraction}, and, respectively, $\mbox{IF}_{\vartheta}^{M}$, for IF based on
{\it reset-to-mod}.

Another concept of level crossing is Send-on-Delta (SOD), which is typically used on continuous signals, see~\cite{Miskowicz2006,Miskowicz2017}.
In this paper we consider also signals with discontinuities. 
Therefore, we define SOD by the mapping 
$\mbox{SOD}_{\vartheta}:\mathcal{F} \rightarrow \mathbb{S}_{\vartheta}$, $s = \mbox{SOD}_{\vartheta}(f)$, $f:[t_a, t_b] \rightarrow \mathbb{R}$, given by
\begin{eqnarray}
    \label{eq:SOD}
    (t_0, s_0) &:=& (t_a, 0) \nonumber \\
    t_{k+1} &:=& \inf\{t>t_k:\, 
    |f(t) - s_k|
    \geq \vartheta\} \nonumber \\
    s_{k+1} &:=& q_{\vartheta}(f(t_{k+1}))-q_{\vartheta}(f(t_{k})),
\end{eqnarray}
which is equivalent with SOD defined on continuous signals~\cite{Miskowicz2006}. Note that with our conception~(\ref{eq:IF}) of IF and the generalization of 
SOD~(\ref{eq:SOD}), 
we obtain the identity
\begin{equation}
\label{eq:IFSOD}
\mbox{SOD}_{\vartheta}{\left(\int_{t_a}^t f(\tau)d\tau\right)} = 
\mbox{IF}_{\vartheta}^{M}{(f)},
\end{equation}
for any integrable $f$ that is bounded except on a finite number of superimposed Dirac pulses.

\section{Mathematical Analysis of Integrate-and-Fire}
\label{sec:math1}
Our mathematical analysis aims at identifying and revealing underlying structural properties of IF as threshold-based sampling scheme.
Later on we will outline such properties (quantization bound, quasi isometry,  maximum sparsity property, characterization as regularization solver). 
It turns out that these properties are only valid under restricted conditions 
for {\it reset-by-subtraction}. In contrast, our recently introduced reset variant {\it reset-to-mod} allows the validity of all these properties also for mixed signals of Dirac-superimposed continuous-time signals.
Under this premise, the mapping~\eqref{eq:IF} induced by IF can be 
reformulated by the following recursion:
\begin{eqnarray}
\label{eq:DefLIFa}
z_{t_1} & := & F_{t_1}, 																	\nonumber \\
s_{t_1} & := & q_{\vartheta}(z_{t_1}), 															\nonumber \\
z_{t_{k+1}} & := & F_{t_{k+1}} + z_{t_k} - s_{t_k},		\nonumber \\
s_{t_{k+1}} & := & q_{\vartheta}(z_{t_{k+1}}),
\end{eqnarray}
where $F_{t_{k+1}}:= \int_{(t_k, t_{k+1}]} f(\tau) d\tau$ is the evaluation of the integral immediately after the spike triggering event at $t = t_k$.
\subsection{IF as Quantization Operator}
\label{ssec:Quantization}
A first result is the following decomposition theorem. For a proof see~\cite{moser2024samplingsparsityArxiv}.
\begin{theorem}[Quantization Operator]
\label{th:quantization}
For any integrable real function $f$ with superposition of finitely many Dirac pulses, i.e., $f$ is of the type $f(t) = \widetilde{f}(t) + \sum_k a_k\delta(t- t_k)$, where $\widetilde{f}:(t_a, t_b] \rightarrow \mathbb{R}$ is bounded and integrable and $t_j \in (t_a, t_b]$, there holds
\begin{equation}
\label{eq:decomp}
\mbox{IF}_{\vartheta}^M(f) = \left(A^{-1} \circ q_{\vartheta} \circ A\right)(f),
\end{equation}
where the standard quantization operation $q$, see~(\ref{eq:q}), 
is applied point-wise and  
the accumulating sum operator $A$ given by $A(f)(t):= \int_0^t f(\tau)d\tau$. 
\end{theorem}

As an immediate corollary of Theorem~(\ref{th:quantization}) we get
\begin{equation}
\label{eq:quant}
\|A\circ [A^{-1} \circ q \circ A (f) - f] \|_{\infty} =
\| \mbox{IF}_{\vartheta}^M(f) - f\|_A < 1,
\end{equation}
which reveals the related metric structure of the underlying signal space, which turns out to be the Alexiewicz semi-norm, 
see~\cite{Alexiewicz1948,MOSER2024128190}, 
\begin{equation}
\label{eq:Alex}
\|f\|_A := \|A(f)\|_{\infty} = \sup_T\left|\int_{t_a}^T f(t)dt\right|.
\end{equation}

Inequality~(\ref{eq:quant}) reflects the behavior of IF as quantization operator in a geometric sense. IF maps the signal $f$ to
the discrete grid-point 
$s = \mbox{IF}_{\vartheta}^M(f)$ in the vector space $\mathbb{S}_{\vartheta} := \{s: s(t) = \sum_j s_j \delta(t - t_j), 
s_j \in \vartheta\mathbb{Z}\}$ of all 
spike trains with discrete times $t_j$ and spike amplitudes being multiples of $\vartheta>0$. By equipping this space with the Alexiewicz norm (now a norm), the points $s\in \mathbb{S}_{\vartheta}$ can be considered as grid points of a tessellation generated by the Alexiewicz ball $\{s \in \mathbb{S}_{\vartheta}: \|s\|_A \leq \vartheta\}$. Then, (\ref{eq:quant}) means that IF/mod maps the signal $f$ to a grid point $s\in \mathbb{S}_{\vartheta}$ that lies strictly within the $\vartheta$-ball $B^A_{\vartheta}(f):= \{g: \|f - g\|_A < \vartheta\}$ centered at $f$.

\subsection{Quasi Isometry}
\label{ssec:QuasiIsometry}
The quantization formula~(\ref{eq:quant}) is fundamental and has various consequences. So, it establishes the quasi-isometry relation, see also~\cite{Moser2017Similarity,MoserLunglmayr2019QuasiIsometry}, 
\begin{equation}
\label{eq:qi}
\|g - f\|_A  - 2 \vartheta < 
\|\mbox{IF}_{\vartheta}^M(g) - \mbox{IF}_{\vartheta}^M(f)\|_A 
< \|g - f\|_A + 2 \vartheta, 
\end{equation}
which immediately follows from the triangle inequality of the semi-norm by considering
\begin{eqnarray}
\left|\,
\|\mbox{IF}_{\vartheta}^M(g) - \mbox{IF}_{\vartheta}^M(f)\|_A
- \|g - f\|_A
\,\right| & \leq & \nonumber \\
\|\mbox{IF}_{\vartheta}^M(g) - g\|_A + \|\mbox{IF}_{\vartheta}^M(f) - f\|_A & < &  2\vartheta. \nonumber 
\end{eqnarray}
While in classical Shannon-based sampling there is an isometry between signal space and the space of its samples, the quasi isometry relation~(\ref{eq:qi}) is the best what can be obtained in the setting of IF-based sampling. The bounds are sharp and global, i.e., independent from any properties of the signals. 
Note that for  thresholds decreasing to zero, we obtain asyomptotic isometry. Further the Alexiewicz norm is uniquely determined up to quasi-isometric norm equivalence to satisfy a quasi-isometry relation
between input and sample space, see ~\cite{Moser2017Similarity,MoserLunglmayr2019QuasiIsometry}.
\subsection{Maximal Sparsity Property}
\label{ssec:sparsity}
For SOD we mention the maximal sampling sparsity that
for any piecewise continuous function $f:[t_a, t_b] \rightarrow \mathbb{R}$ with $f(t_a)=0$, SOD-based threshold-based sampling with threshold $\vartheta>0$ satisfies the property
\begin{equation}
\label{eq:SODsparse}
\|\mbox{SOD}_{\vartheta}(f)\|_1 = 
 \min\{\|s\|_1:  
\| f - z_s \|_{\infty}<\vartheta,
z_s, s \in \mathbb{S}_{\vartheta}\}, 
\end{equation}
where
$z_s(t):= \int_{t_a}^t\sum_{t_k \leq t} s_k\, \delta(\tau - t_k) d\tau$ for 
$s \in \mathbb{S}_{\vartheta}$ given by $s(t)=\sum_{t_k \leq t} s_k\, \delta(\tau - t_k)$.
An analogous characterization can be proven for IF/mod. 
For the proof we refer to a paper currently under review, see~\cite{moser2024samplingsparsityArxiv}.
For IF/mod, this property states that $s = \mbox{IF}_{\vartheta}^M(f)$ is not some arbitrary point within the $\vartheta$-Alexiewicz ball centered at $f$, rather it is a uniquely determined grid point, i.e., spike train, that distinguishes by its minimal $l_1$-norm, i.e., maximal sparsity.
\begin{theorem}[Maximal Sparsity Property of IF/mod]
\label{th:sparsenessIF}
For any function $f:(t_a, t_b] \rightarrow \mathbb{R}$  that is integrable and almost everywhere bounded with locally finite many superimposed Dirac impulses, 
IF with threshold $\vartheta>0$ satisfies the maximal sparsity property
\begin{equation}
\label{eq:sparsenessProp}
\|\mbox{IF}_{\vartheta}^M(f)\|_1 = \min\{\|s\|_1:  s \in \mathring{B}^A_{\vartheta}(f) \cap \mathbb{S}_{\vartheta}  \},
\end{equation}
where $\mathring{B}^A_{\vartheta}(f):= \{g: \|g - f\|_A < \vartheta\}$ 
represents the open Alexiewicz ball with radius $\vartheta$ centered at $f$.
\end{theorem}

\section{Signal Reconstruction}
\label{s:Reconstruction}
First, we consider the generalized version of SOD~(\ref{eq:SOD}), which can also be applied to bounded piecewise continuous functions $g: [t_a, t_b] \rightarrow \mathbb{R}$. We assume to start the process at $t = t_a$ with $g(t_a)=0$.
Let $s = \mbox{SOD}_{\vartheta}(g)$, i.e., 
$s(t) = \sum_k s_k \delta(t - t_k)$.
The spikes $(t_k, s_k)$ make it possible to construct a step function $\overline{g}_{\mbox{\tiny{SOD}}}$
such that 
\begin{equation}
\| g - \overline{g}_{\mbox{\tiny{SOD}}} \|_{\infty} < \vartheta,    
\end{equation}
where
\begin{equation}
\label{eq:SODrec0}
    \overline{g}_{\mbox{\tiny{SOD}}}(t) := 
 \sum_k a_k 1_{[t_k, t_{k+1})}(t), a_k = \sum_{j\leq k} s_j
\end{equation}
and $1_{I}(t)=1$ if $t\in I$ otherwise $1_{I}(t)=0$.
As alternative one might prefer a piecewise linear function $\widetilde{g}_{\mbox{\tiny{SOD}}}$
such that 
\[
\| g - \widetilde{g}_{\mbox{\tiny{SOD}}} \|_{\infty} < 2 \vartheta,
\]
where
\begin{equation}
\label{eq:SODrec}
    \widetilde{g}_{\mbox{\tiny{SOD}}}(t) := 
 \sum_k 1_{[t_k, t_{k+1})}(t)\, G_k(t),
\end{equation}
\[
G_k(t) := 
\sum_{j \leq k} s_j + 
1_{\{|s_{k+1}| \leq \vartheta\}}
(s_{k+1})\,\vartheta\frac{\mbox{sgn}(s_{k+1})}{t_{k+1} - t_k}(t -  t_k).
\]
For IF we consider the wider class of functions $f$ that are bounded and integrable with a superposition of finitely many Dirac impulses. With the notation $z := \mbox{IF}_{\vartheta}^M(f)$,  $z(t) = \sum_k z_k\, \delta(t - t_k)$, we obtain the following formula for a reconstructions based on {\it reset-to-mod}.
Analogous to~(\ref{eq:SODrec0}) we obtain 
\begin{equation}
\label{eq:IFrec0}
\overline{f}_{\mbox{\tiny{IF/mod}}}(t) := 
 \sum_k z_k \delta(t-t_k),
\end{equation}
where $(t_k, z_k)$ are the spikes resulting from 
$\mbox{IF}_{\vartheta}^M(f)$.
And, analogous to (\ref{eq:SODrec}) we obtain
\begin{equation}
\label{eq:IFrec}
\widetilde{f}^M(t) := 
 \sum_k 1_{[t_k, t_{k+1})}(t)\, F_k^M(t),
\end{equation}
\begin{eqnarray}
    F_k^M(t) &:= &
1_{\{|z_k|>\vartheta\}}(z_k)\delta(t-t_k)q_{\vartheta}(z_k) + \nonumber \\
 & &  1_{\{|z_{k+1}|\leq \vartheta\}}(z_{k+1})
\frac{\vartheta\,\mbox{sgn}(z_{k+1})}{t_{k+1} - t_k}. \nonumber
\end{eqnarray}
Since for IF based on {\it reset-by-subtraction} the spike amplitudes $z_k$ are restricted to 
$|z_k| \leq \vartheta$, 
(\ref{eq:IFrec}) reduces to the simpler form
\begin{equation}
\label{eq:IFSubrec}
\widetilde{f}^S(t) := 
 \sum_k 1_{[t_k, t_{k+1})}(t)\, \frac{\vartheta\,\mbox{sgn}(z_{k+1})}{t_{k+1} - t_k}.
 \end{equation}

\begin{theorem}[IF/mod-Reconstruction]
\label{th:Reconstruction}
The functions $\widetilde{g}_{\mbox{\tiny{SOD}}}$ and $\widetilde{f}^M$ given by (\ref{eq:SODrec}) and (\ref{eq:IFrec}), respectively, satisfy the properties: (i) $\mbox{SOD}_{\vartheta}(g) = \mbox{SOD}_{\vartheta}(\widetilde{g}_{\mbox{\tiny{SOD}}})$, (ii) $\mbox{IF}_{\vartheta}^M(f) = \mbox{IF}^M_{\vartheta}(\widetilde{f}^M)$, and (iii) 
$\widetilde{g}_{\mbox{\tiny{SOD}}}(t) := 
\int_{t_a}^t \widetilde{f}^M(\tau)d\tau$ gives a reconstruction w.r.t the spike train obtained from $\mbox{SOD}_{\vartheta}$ applied on $\int_{t_a}^t f(\tau)d\tau$.
\end{theorem}
If no Dirac impulses occur, (\ref{eq:IFrec}) becomes continuous, yielding the standard realization of a step function with the step amplitudes $z_{k+1}$ as integral of $f$ over $[t_k, t_{k+1}]$. 

\section{Sparse Regularization}
\label{s:TV}
\begin{figure}[htbp]
\centering
\includegraphics[width=0.7\linewidth]{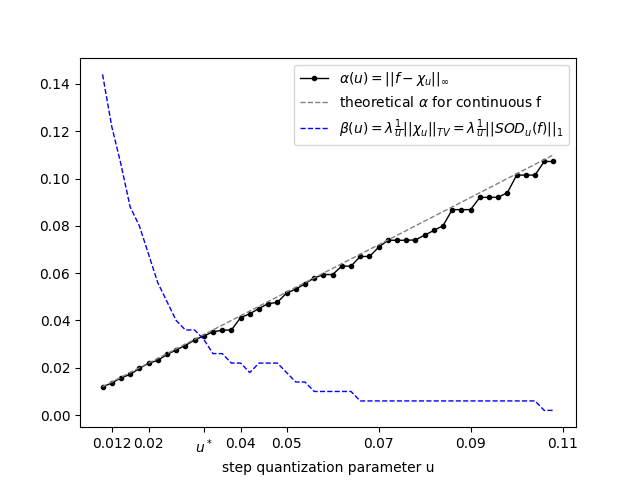}
\caption{Illustration of (\ref{eq:TVreg})  with typical curves $\alpha$, $\beta$ of opposite monotonicity behavior. $u$ denotes the quantization step in (\ref{eq:TVreg}). The simulation is based on the velocity data of Fig.~\ref{fig:IFVel02} with  $\lambda=0.002$. 
As the data is discrete, $\alpha(u)=u$ is only approximately valid.
}
\label{fig:regularization}
\end{figure}

Consider the space $\mathcal{U}$ of step functions with quantized steps, i.e., $\chi_u = \sum_{k=1}^N u\,c_k 1_{I_k} \in \mathcal{U}$ for some step quantization $u>0$ and $n_k \in, \mathbb{Z}$ for all $k$. $\chi_u$ is sparsely encoded if $N$ and the step differences are low. 
This way, we get $\frac{1}{u}\|\chi_u\|_{TV}$ as sparsity measure of $\chi_u$. 
Then, the sparsity property~(\ref{eq:SODsparse}) entails 
the following characterization as sparse regularization in terms of total variation (TV). 
\begin{theorem}[SOD as sparse-regularization solver]
Let $\lambda>0$.
For any continuous $f$ with $\|f\|_{TV}>0$ the sparse regularization problem
    \begin{equation}
    \label{eq:TVreg}
    \min_{\chi_u \in \mathcal{U}} \|f - \chi_u\|_{\infty} + \lambda \frac{1}{u}\|\chi_u\|_{TV}
\end{equation}
is solved by $s_{\vartheta} := \mbox{SOD}_{\vartheta}(f)$, $s_{\vartheta}(t) = 
\sum_k s_k \delta(t - t_k)$, and its induced step function 
$\chi_{\vartheta}(t) := \sum_k a_k 1_{[t_k, t_{k+1})}(t)$,  
$a_k = \sum_{j\leq k} s_j$, where $\vartheta=u^*$ is determined by the unique intersection point $u^* = \|\mbox{SOD}_{u^*}(f)\|_1$.
\end{theorem}
The proof relies on $\alpha(u):= \|f - \chi_u\|_{\infty}=u$ for continuous $f$ and the convexity of $\beta(u):=\frac{1}{u}\|\chi_u\|_{\mbox{\tiny{TV}}}$, where $\chi_u$ is defined by (\ref{eq:SODrec0}). See Fig.~\ref{fig:regularization} for an illustration. An extended proof is postponed to an upcoming paper. 
Vice versa, for a given $\vartheta$ one can find a $\lambda$ such that we obtain the SOD-induced step function (\ref{eq:SODrec0}) as solver of the regularization problem~(\ref{eq:TVreg}). 

In an analogous way, we obtain a characterization of IF/mod in terms of regularizing the Alexiewicz norm. 
For a piecewise continuous $f$ with 
$\|f\|_{\mbox{TV}}>0$ the regularization problem 
\[
\min_{\psi_u} \|f - \psi_u\|_A + \lambda\, \frac{1}{u}\|\psi_u\|_1,
\]
among all impulse sequences $\psi_u(t) = \sum_k u\,b_k \delta(t-t_k)$ with $b_k\in \mathbb{Z}$,
is solved by $\psi_{\vartheta} := \mbox{IF}^{M}_{\vartheta}(f)$,
where $\vartheta:=u^*$ is determined as unique intersection point of $u^* = \|\mbox{IF}^{M}_{u^*}(f)\|_1$.

\section{Examples and Evaluations}
\label{s:Examples}
\begin{figure}[htbp]
\centering
\includegraphics[width=0.7\linewidth]{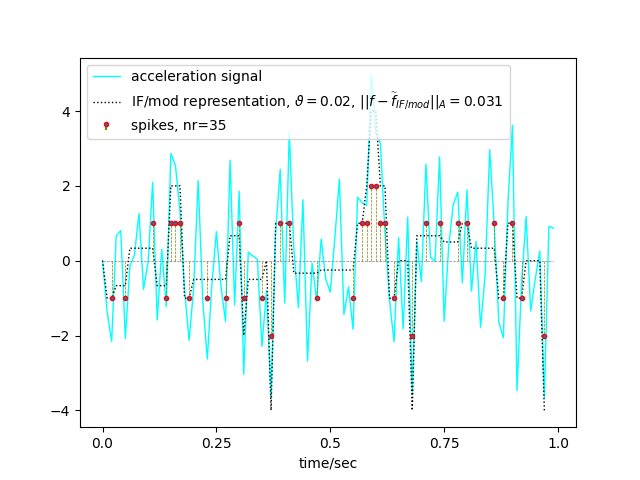}
\caption{$100$Hz acceleration data and its sparse representation by spikes obtained by integrate-and-fire (\ref{eq:IF}) based on {\it reset-to-mod} and its induced reconstruction~\ref{eq:IFrec}). Here, spike amplitudes are multiples of the threshold. Due to~(\ref{eq:quant}) and~(\ref{eq:IFrec}) we have $\|f  -  \mbox{IF}^M_{\vartheta}(f)\|_A  \leq  \vartheta$ and
$\|f  -  \widetilde{f}^M\|_A \leq 2\vartheta$.}
\label{fig:IFmodAcc2}
\end{figure}

\begin{figure}[htbp]
\centering
\includegraphics[width=0.7\linewidth]{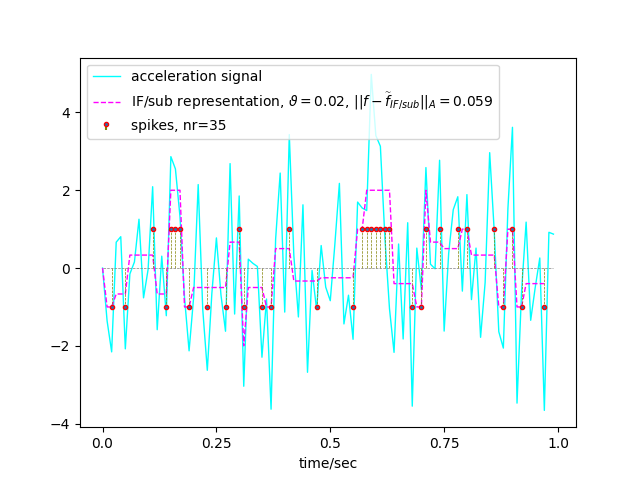}
\caption{Like Fig.~\ref{fig:IFmodAcc2}, but now with IF based on {\it reset-by-subtraction} and its reconstruction~\ref{eq:IFSubrec}). Here, spike amplitudes are restricted to $\pm \vartheta$.  $\|f  -  \widetilde{f}_{\mbox{\tiny{IF/sub}}}\|_A \leq \vartheta$ is not valid any more in general.}
\label{fig:IFsubAcc2}
\end{figure}

\begin{figure}[htbp]
\centering
\includegraphics[width=0.7\linewidth]{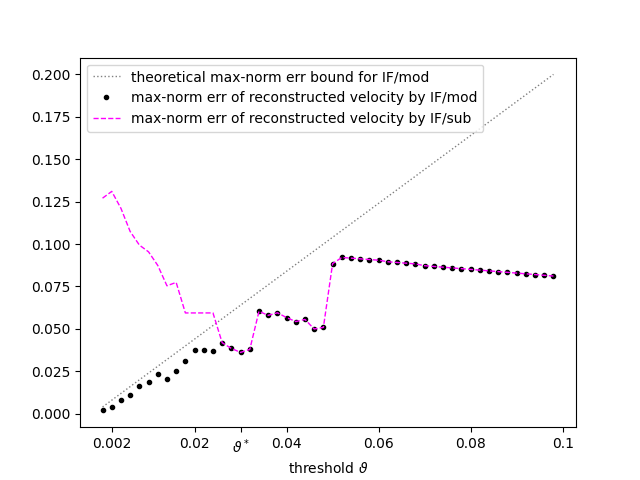}
\caption{$\|.\|_{\infty}$-error between velocity and its IF-based reconstructions for $\vartheta \in (0, 0.1)$. Due to~(\ref{eq:quant}) and~(\ref{eq:IFrec}), for IF/mod  the error is $O(\vartheta)$, while IF/sub variant becomes instable for too small $\vartheta$. $\vartheta^*$ denotes the threshold with minimal max-norm reconstruction error based on IF/sub.  
}
\label{fig:IFerrConvergence}
\end{figure}

We carry out experiments using acceleration data (taken from ~\cite{Sampaio2019}) to illustrate the formulas 
(\ref{eq:quant}), (\ref{eq:SODrec}), (\ref{eq:IFrec})  and the difference between the state of the art of integrate-and-fire using {\it reset-by-subtraction}, see Fig.~\ref{fig:IFsubAcc2}, and our modification using {\it reset-to-mod}, see Fig.\ref{fig:IFmodAcc2}. Fig.~\ref{fig:IFerrConvergence} outlines the related behavior of the max-norm error for both variants for small thresholds $\vartheta\rightarrow 0$. The different behavior of IF based on {\it reset-to-mod} compared to {\it reset-by-subtraction} becomes more evident at small threshold values in relation to the signal slopes.
Since the max-error of IF/sub increases for small and large thresholds, there is an optimal threshold $\vartheta^*$.
For $\vartheta^*$, the corresponding reconstruction provides a continuous piecewise function with error bound below $2\vartheta^*$.
For sufficiently large threshold values both reset variants behave approximately the same. Let $f_i$ be the equidistant samples of $f$ at sampling rate $1/\Delta T$. 
Then we obtain as a sufficient condition that the reconstructions  by (\ref{eq:IFrec}) and (\ref{eq:IFSubrec}) are identical, i.e., 
$\widetilde{f}^{S}(t) = \widetilde{f}^M(t)$ if $\max_i |f_i|\, \Delta T < \vartheta$.

\begin{figure}[htbp]
\centering
\includegraphics[width=0.7\linewidth]{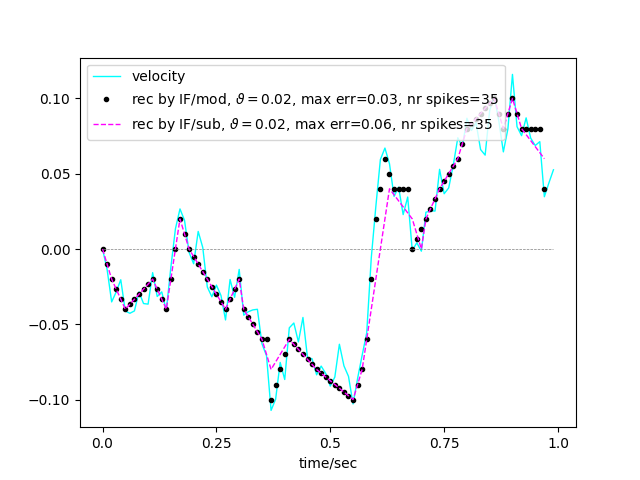}
\caption{Velocity data obtained from integrating the acceleration data in Fig.~\ref{fig:IFmodAcc2}, resp. Fig.~\ref{fig:IFsubAcc2} and its IF-based reconstructions, comparing the {\it reset-to-mod} and the {\it reset-by-subtraction} variants.}
\label{fig:IFVel02}
\end{figure}


\section{Conclusion}
\label{s:Conclusion}
We showed that integrate-and-fire (IF) based on {\it reset-to-mod} can be considered as integration-based version of SOD, operating on signals that are bounded and integrable except a discrete superposition of Dirac impulses. As with SOD, IF also allows for signal reconstruction. However, the related error bounds require the consideration of different metrics. So, the counterpart of the max norm for SOD turns out to be the Alexiewicz norm for IF. 
SOD and IF are sampling schemes that operate in the range-domain which equivalently can be understood 
from a regularization perspective. In future research, we will exploit this connection to regularization theory. Further, we will exploit the generalized SOD for a new design approach to realize new SOD analog-to-spike converters for high-frequency applications.

\end{document}